\documentclass[conference,letterpaper]{IEEEtran}

\usepackage[utf8]{inputenc} 
\usepackage[T1]{fontenc}
\usepackage{graphicx}
\usepackage{url}
\usepackage{ifthen}
\usepackage{cite}
\usepackage{amssymb}
\usepackage{bm}
\usepackage[cmex10]{amsmath} 

\newtheorem{theorem}{Theorem}

\begin{document}
\title{Improved Computation-Communication Trade-Off for Coded Distributed Computing using Linear Dependence of Intermediate Values} 

 \author{%
   \IEEEauthorblockN{Shunsuke Horii}
   \IEEEauthorblockA{Waseda University\\
                     1-6-1, Nishiwaseda, Shinjuku-ku,\\
                     Tokyo 169-8050, Japan\\
                     Email: s.horii@aoni.waseda.jp}
 }

\maketitle

\begin{abstract}
   In large scale distributed computing systems, communication overhead is one of the major bottlenecks.
   In the map-shuffle-reduce framework, which is one of the major distributed computing frameworks, the communication load among servers can be reduced by increasing the computation load of each server, that is, there is a trade-off between computation load and communication load.
   Recently, it has been shown that coded distributed computing (CDC) improves this trade-off relationship by letting servers encode their intermediate computation results.
   The original CDC scheme does not assume any special structures on the functions that servers compute.
   However, in actual problems, these functions often have some structures, and the trade-off relation may be further improved by using that structures.
   In this paper, we propose a new scheme that further improves the trade-off relationship by utilizing the linear dependency structure of the intermediate computation results.
   The intermediate values computed in the map phase can be considered as vectors on $\mathbb{F}_{2}$.
   In some applications, these intermediate values have a linear dependency and in such cases, it is sufficient for each server to send a basis of the linear subspace and linear combination coefficients.
   As a result, the proposed approach improves over the best-known computation-communication overhead trade-off in some applications.
\end{abstract}


\section{Introduction}

The development in measurement technology and network technology leads to a reduction in data collection costs.
As a result, the amount of data processed by computers has been increasing dramatically, and it becomes difficult to process data by a single computer or a single processor.
Therefore, the distributed computing system, in which data is distributed to many computers or processors and processed in parallel, is widely used.

There are two main bottlenecks in distributed computing systems.
One is the existence of straggling servers.
Servers that take a long time to complete their tasks or to communicate with other servers for some problems are called stragglers.
Servers that for some reason take a long time to complete their tasks or have trouble communicating with other servers are called stragglers.
If no countermeasures are taken, the computation time of the entire system depends on the computation time of the server with the slowest computation time, so the existence of straggler increases the overall computation time.
Recently, there are many studies to reduce the effect of the stragglers by using the error-correcting codes \cite{lee2018speeding,dutta2016short,yu2017polynomial,karakus2017straggler,tandon2017gradient,li2018near,ye2018communication,halbawi2018improving,8849580}.
Error-correcting codes enable distributed computing systems to obtain the computation result even if there are some stragglers.
Another bottleneck is the communication overhead in data shuffling among the servers.
As the number of servers increases, the communication time to share the intermediate computation results would increase.
Increasing the load of the local computation of servers can reduce the communication load, but then the local computation time on each server would be dominant, so there is a trade-off between the computation load and the communication load.
Recently, coded distributed computing (CDC) has been proposed in \cite{li2017fundamental} to improve this trade-off by letting servers encode their intermediate computation results.
We also note that some studies combine two coding schemes, that is, CDC that is resilient to stragglers \cite{li2016unified,zhang2019improved}.

In this paper, as in \cite{li2017fundamental,li2016unified,zhang2019improved}, we consider the map-shuffle-reduce framework, which is one of the major distributed computing frameworks.
As the name suggests, the framework consists of three phases: map, shuffle and reduce.
In the map phase, each server processes some map functions and outputs intermediate computation results (intermediate values).
In the shuffle phase, servers exchange intermediate values by communicating with each other.
Finally, servers compute some reduce functions to obtain the final output result.
In the map phase, if servers compute all necessary intermediate values, no communication is required.
On the other hand, if servers compute some part of the necessary intermediate values, they have to obtain the rest of them through their communication.
Thus, there is a trade-off relationship between the computation load and the communication load.

Without any coding scheme, each server sends intermediate values without any processing.
In \cite{li2017fundamental}, it has shown that CDC scheme improves the computation-communication trade-off.
In the CDC scheme, each server broadcasts some linear combinations of some part of the intermediate values so that other servers can recover the necessary intermediate values.
In \cite{li2017fundamental}, it is also shown that the proposed CDC scheme is tight, that is, no scheme has better computation-communication trade-off \textit{without any further assumptions}.
However, in some applications, the functions that servers compute often have some structures and the computation-communication trade-off may be further improved by using that structures.
Actually, in \cite{li2018compressed}, it proposes a new scheme which has better computation-communication trade-off compared to the original one for the problems that the reduce functions are linear aggregation.

In the CDC scheme, the encoded messages are represented as elements of $\mathbb{F}_{2^{\ell}}$ for some $\ell$.
Thus, each server sends some elements of $\mathbb{F}_{2^{\ell}}$.
We can also regard elements of $\mathbb{F}_{2^{\ell}}$ as length $\ell$ vectors of $\mathbb{F}_{2}$, so the information sent by each server can be regarded as linear subspace.
Therefore, even if the number of elements sent by a server is $r$, the dimension of linear subspace constructed from those vectors may be lower than $r$.
For example, if the problem is to count the number of words in large amounts of documents, many words do not appear in some parts of documents.
In such cases, many intermediate values computed by a server take same values and some of their linear combinations are also same.
Then the dimension of linear subspace constructed from coded symbols is smaller than the number of coded symbols.
It is sufficient for each server to send the basis of the linear subspace and linear combination coefficients.
This result leads to the further improvement of the computation-communication trade-off.

The rest of the paper is organized as follows.
In Section 2, we describe system model and basic background about coding scheme for map-shuffle-reduce framework.
In Section 3, we establish our main results with a motivating example.
Finally, we give a summary and future works in Section 4.

\section{System model and background}

In this section, we illustrate the map-shuffle-reduce framework for distributed computing and CDC scheme.
We will follow along the same line established in \cite{li2017fundamental}.
The decoding scheme is not closely related to our research, so this paper described only the encoding scheme and does not explain the decoding scheme.
See \cite{li2017fundamental} for more detail.

\subsection{System Model}
 
We consider the problem of computing $Q$ output functions from $N$ input files.
Given $N$ input files $w_{1},\ldots,w_{N}\in\mathbb{F}_{2^{F}}$, for some $F\in\mathbb{N}$, the goal is to compute $Q$ output functions $\phi_{1},\ldots,\phi_{Q}$, where $\phi_{q}:\left(\mathbb{F}_{2^{F}}\right)^{N}\to \mathbb{F}_{2^{B}}, q\in\left\{1,\ldots,Q\right\}$.
We assume that the output functions $\phi_{q},q\in\left\{1,\ldots,Q\right\}$ can be decomposed as follows:
\begin{align}
    \phi_{q}(w_{1},\ldots,w_{N})=h_{q}(g_{q,1}(w_{1}),\ldots,g_{q,N}(w_{N})),
\end{align}
where
\begin{itemize}
    \item The "Map" function $g_{q,n}:\mathbb{F}_{2^{F}}\to \mathbb{F}_{2^{T}}$ maps the input file $w_{n}$ into the length-$T$ intermediate value for some $T\in\mathbb{N}$. The intermediate value, which is the output of the map function $g_{q,n}$, is denoted by $v_{q,n}$, i.e., $v_{q,n}=g_{q,n}(w_{n})$.
    \item The "Reduce" function $h_{q}:\left(\mathbb{F}_{2^{T}}\right)^{N}\to \mathbb{F}_{2^{B}}$ maps the intermediate values of the map functions into the output value.
\end{itemize}

The computation of map and reduce functions are carried out by $K$ distributed computing nodes (servers).
They are interconnected through an error-free broadcast network.

\subsection{Coded Map-Shuffle-Reduce Framework}
Node $k$ computes the map functions of a set of files $\mathcal{M}_{k}\subseteq\left\{w_{1},\ldots,w_{N}\right\}$.
The computation load is defined as $r\triangleq \frac{\sum_{k=1}^{K}|\mathcal{M}_{k}|}{N}$.
For the sake of simplicity, we assume that the number of input files $N$ can be divided by $\binom{K}{r}$.
In the map phase, the $N$ input files are evenly partitioned into $\binom{K}{r}$ disjoint batches of size $\eta_{1}=N/\binom{K}{r}$, each corresponding to a subset $\mathcal{T}\subset \left\{1,\ldots,K\right\}$ of size $r$, i.e., 
\begin{align}
\left\{w_{1},\ldots,w_{N}\right\}=\cup_{\mathcal{T}\subset\left\{1,\ldots,K\right\},|\mathcal{T}|=r}\mathcal{B}_{\mathcal{T}}
\end{align}
where $\mathcal{B}_{\mathcal{T}}$ denotes the batch of $\eta_{1}$ files corresponding to the subset $\mathcal{T}$.
Node $k$ computes the map functions of the files in $\mathcal{B}_{\mathcal{T}}$ iff $k\in\mathcal{T}$.

\begin{figure}[t]
   \centering
   \includegraphics[keepaspectratio=true, width=\linewidth]{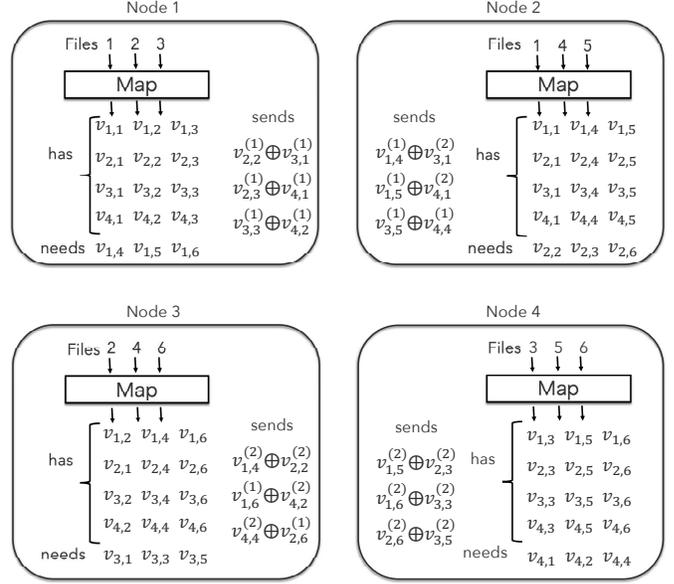}
   \caption{An example of the coded map-shuffle-reduce system. $v_{q,n}^{(1)}$ and $v_{q,n}^{(2)}$ are first and second half segments of $v_{q,n}$, respectively. For example, node 1 can recover $v_{1,4}^{(1)}$ by subtracting $v_{3,1}^{(2)}$ from $v_{1,4}^{(1)}\oplus v_{3,1}^{(2)}$ sent by node 2 and recover $v_{1,4}^{(2)}$ by subtracting $v_{2,2}^{(2)}$ from $v_{1,4}^{(2)}\oplus v_{2,2}^{(2)}$ sent by node 3, thus it can recover $v_{1,4}$.} 
   \label{fig:example}
\end{figure}

Node $k$ computes a subset of output functions, whose indices are denoted by $\mathcal{W}_{k}\subseteq\left\{1,\ldots,Q\right\}$.
Again, for the sake of simplicity, we assume that the number of the reduce functions $Q$ can be divided by $\binom{K}{s}$, and every subset of $s$ nodes reduce $Q/\binom{K}{s}$ functions.
The parameter $s$ is the number of times each reduce function is computed.
The $Q$ reduce functions are evenly partitioned into $\binom{K}{s}$ disjoint batches of size $\eta_{2}=Q/\binom{K}{s}$, each corresponding to a subset $\mathcal{P}$ of $s$ nodes, i.e.,
\begin{align}
    \left\{1,\ldots,Q\right\}=\cup_{\mathcal{P}\subset\left\{1,\ldots,K\right\},|\mathcal{P}|=s}\mathcal{D}_{\mathcal{P}},
\end{align}
where $\mathcal{D}_{\mathcal{P}}$ denotes the indices of the batch of $\eta_{2}$ reduce functions corresponding to the subset $\mathcal{P}$.
Node $k$ computes the reduce functions in the set $\mathcal{D}_{\mathcal{P}}$ iff $k\in\mathcal{P}$.
An example of the distributed computing system is depicted in Fig. \ref{fig:example} for $K=4, N=6, Q=4, r=2, s=1$.
For example, in this example, $\mathcal{B}_{\left\{1,2\right\}}=\left\{1\right\}$ and $\mathcal{D}_{\left\{1\right\}}=\left\{1\right\}$.

If the nodes are not allowed to utilize any coding scheme, each node has to receive the necessary intermediate values sent without coding by some other nodes.
The communication load (the precise definition is given later) achieved by the uncoded scheme is 
\begin{align}
    L_{\mbox{uncoded}}(r)=1-r/K.
\end{align}

In the CDC scheme, the nodes construct the coded messages as follows.
Let $\mathcal{S}$ be a subset of $\left\{1,\ldots,K\right\}$ of size $\max\left\{r+1,s\right\}\le |\mathcal{S}|\le\min\left\{r+s, K\right\}$.
For a subset $\tilde{S}\subset \mathcal{S}$ with $|\tilde{\mathcal{S}}|=r$, let $\mathcal{V}_{\tilde{\mathcal{S}}}^{\mathcal{S}\setminus \tilde{\mathcal{S}}}$ be the set of intermediate values needed by all nodes in $\mathcal{S}\setminus \tilde{\mathcal{S}}$, not required by nodes outside $\mathcal{S}$, and known exclusively by nodes in $\tilde{\mathcal{S}}$, i.e., 
\begin{multline}
    \mathcal{V}_{\tilde{\mathcal{S}}}^{\mathcal{S}\setminus\tilde{\mathcal{S}}}=\left\{v_{q,n}:q\in\cap_{k\in\mathcal{S}\setminus \tilde{\mathcal{S}}}\mathcal{W}_{k},q\notin\cup_{k\notin\mathcal{S}}\mathcal{W}_{k}\right.,\\
    \left.w_{n}\in\cap_{k\in\tilde{\mathcal{S}}}\mathcal{M}_{k},w_{n}\notin\cup_{k\notin\tilde{\mathcal{S}}}\mathcal{M}_{k}\right\}.
\end{multline}
For example, $\mathcal{V}^{\left\{2\right\}}_{\left\{1,3\right\}}=\left\{v_{2,2}\right\}$ for the example in Fig. \ref{fig:example}.
The set $\mathcal{V}_{\tilde{\mathcal{S}}}^{\mathcal{S}\setminus \tilde{\mathcal{S}}}$ contains $\binom{r}{|\mathcal{S}|-s}\eta_{1}\eta_{2}$ intermediate values.
A symbol $U_{\tilde{\mathcal{S}}}^{\mathcal{S}\setminus \tilde{\mathcal{S}}}\in\mathbb{F}_{2^{\binom{r}{|\mathcal{S}|-s}\eta_{1}\eta_{2}T}}$ is the concatenation of the intermediate values in $\mathcal{V}_{\tilde{\mathcal{S}}}^{\mathcal{S}\setminus \tilde{\mathcal{S}}}$.
For $\tilde{\mathcal{S}}=\left\{\sigma_{1},\ldots,\sigma_{r}\right\}$, $U_{\tilde{\mathcal{S}}}^{\mathcal{S}\setminus \tilde{\mathcal{S}}}$ is split into $r$ segments, each containing $\binom{r}{|\mathcal{S}|-s}\frac{\eta_{1}\eta_{2}T}{r}$ bits, i.e., 
\begin{align}
    U_{\tilde{\mathcal{S}}}^{\mathcal{S}\setminus \tilde{\mathcal{S}}}=\left( U_{\tilde{\mathcal{S}},\sigma_{1}}^{\mathcal{S}\setminus \tilde{\mathcal{S}}},\ldots, U_{\tilde{\mathcal{S}},\sigma_{r}}^{\mathcal{S}\setminus \tilde{\mathcal{S}}}\right).
\end{align}
r example, $U^{\left\{2\right\}}_{\left\{1,3\right\},1}=v_{2,2}^{(1)}$ and $U^{\left\{2\right\}}_{\left\{1,3\right\},3}=v_{2,2}^{(2)}$, where $v_{2,2}^{(1)}$ and $v_{2,2}^{(1)}$ are the first half and second half bits of $v_{2,2}$, respectively.
The node $\sigma_{i}\in\tilde{\mathcal{S}}$ is responsible for $U_{\tilde{\mathcal{S}},\sigma_{i}}^{\mathcal{S}\setminus \tilde{\mathcal{S}}}$.
For each $k\in\mathcal{S}$, there are a total of $m_{\mathcal{S}}=\binom{|\mathcal{S}|-1}{r-1}$ subsets of $\mathcal{S}$ with size $r$ that contain the node $k$.
We index these subsets as $\mathcal{S}_{(k)}[1],\ldots,\mathcal{S}_{(k)}[m_{\mathcal{S}}]$.
Let $n_{\mathcal{S}}=\binom{|\mathcal{S}|-2}{r-1}$, then the the coded messages $\bm{X}_{k}^{\mathcal{S}}=(X_{k}^{\mathcal{S}}[1],\ldots,X_{k}^{\mathcal{S}}[n_{\mathcal{S}}])$, which are broadcast by the node $k$ to the nodes in $\mathcal{S}$ is constructed as follows.
\begin{align}
    \begin{bmatrix}
    X_{k}^{\mathcal{S}}[1]\\X_{k}^{\mathcal{S}}[2]\\ \vdots\\ X_{k}^{\mathcal{S}}[n_{\mathcal{S}}]
    \end{bmatrix}
    &=A_{\mathcal{S}}
    \begin{bmatrix}
    U_{\mathcal{S}_{(k)}[1],k}^{\mathcal{S}\setminus \mathcal{S}_{(k)}[1]}\\
     U_{\mathcal{S}_{(k)}[2],k}^{\mathcal{S}\setminus \mathcal{S}_{(k)}[2]}\\
     \vdots\\
      U_{\mathcal{S}_{(k)}[m_{\mathcal{S}}],k}^{\mathcal{S}\setminus \mathcal{S}_{(k)}[m_{\mathcal{S}}]}
    \end{bmatrix},\label{X_k_S}\\
    A_{\mathcal{S}}&=\begin{bmatrix}
    1 & 1 & \ldots & 1\\
    a_{1} & a_{2} & \ldots & a_{m_{\mathcal{S}}}\\
    \vdots & \vdots & \ddots & \vdots\\
    a_{1}^{n_{\mathcal{S}}-1} & a_{2}^{n_{\mathcal{S}}-1} & \ldots & a_{m_{\mathcal{S}}}^{n_{\mathcal{S}}-1}
    \end{bmatrix},
\end{align}
where coefficients $a_{1},\ldots,a_{m_{\mathcal{S}}}\in \mathbb{F}_{2}^{\binom{r}{|\mathcal{S}|-s}}\frac{\eta_{1}\eta_{2}T}{r}$ are designed such that the nodes in $\mathcal{S}$ can decode the messages $\bm{X}_{k}^{\mathcal{S}}$.
See \cite{li2017fundamental} for detailed conditions.
An example of the constructed message is depicted in Fig. \ref{fig:example}.
When $s=1$, only subsets of $\left\{1,\ldots,K\right\}$ of size $r+1$ are chosen as $\mathcal{S}$ and $A_{\mathcal{S}}=\left[1\ 1\ \ldots\ 1\right]$ for all $\mathcal{S}$, so the encoded messages are the XORs of some parts of the intermediate values.

The communication load is defined as $L\triangleq \frac{\sum_{k=1}^{K}b_{k}}{QNT}$, where $b_{k}$ is the number of bits sent by the node $k$.
In \cite{li2017fundamental}, it is shown that the communication load of the coded scheme described above is 
\begin{align}
    L_{\mbox{CDC}}^{*}(r,s)=\sum_{\ell=\max\left\{r+1,s\right\}}^{\min\left\{r+s,K\right\}}\frac{\ell\binom{K}{\ell}\binom{\ell-2}{r-1}\binom{r}{\ell-s}}{r\binom{K}{r}\binom{K}{s}}.
\end{align}
Especially, when $s=1$ (each reduce function is computed only once by a node), the communication load is given by
\begin{align}
    L_{\mbox{CDC}}^{*}(r)=\frac{1}{r}\left(1-\frac{r}{K}\right).
\end{align}

In \cite{li2017fundamental}, it is also shown that no scheme has smaller communication load for the same computation load if there is no further assumptions.
However, as stated in the introduction, the map functions or reduce functions often have some structures in some applications.
In such cases, the communication load may be further improved by utilizing these structures.
In \cite{li2018compressed}, it shows that the scheme that combines CDC scheme and compression scheme has better computation-communication trade-off when the reduce functions are linear aggregation, so this method utilizes the structure of the reduce functions.
On the other hand, the method proposed in this paper utilizes the structure of the map functions.

\section{Coded map-shuffle-reduce scheme using linear dependency}\label{section2}

\subsection{Illustrative Example}

Here, we present an illustrative example of the proposed scheme.

Consider a problem to counter the number of appearance of the numbers in the following sequence.
\[
    \underbrace{1212231}_{w_{1}}\ \underbrace{2111121}_{w_{2}}\ \underbrace{2312131}_{w_{3}}\ \underbrace{3112132}_{w_{4}}\ \underbrace{1131414}_{w_{5}}\ \underbrace{1141231}_{w_{6}}
\]
The sequence has 4 numbers.
We consider a distributed computing system with $K=4, N=6$, where the input files $w_{1},\ldots,w_{6}$ are the 6 blocks of the sequence each of length 7.
We assume that the node $k$ counts the number of appearance of the number $k$.
Then, the intermediate value $v_{q,n}$ represents the number of '$q$'s in the $n$-th block of the sequence.
For example, $v_{1,1}=3, v_{1,2}=5, v_{1,3}=3$.
Consider the CDC scheme presented in the previous section with $r=2$, which is the same as the example presented in Fig. \ref{fig:example}.
Then, the node broadcasts $v_{2,2}^{(1)}\oplus v_{3,1}^{(1)}$, $v_{2,3}^{(1)}\oplus v_{4,1}^{(1)}$ and $v_{3,3}^{(1)}\oplus v_{4,2}^{(1)}$.
Since $v_{2,3}=v_{3,3}=2$ and $v_{4,1}=v_{4,2}=0$, it holds $v_{2,3}\oplus v_{4,1}=v_{3,3}\oplus v_{4,2}$.
As a consequence, it also holds $v_{2,3}^{(1)}\oplus v_{4,1}^{(1)}=v_{3,3}^{(1)}\oplus v_{4,2}^{(1)}$.
This fact indicates that the rank of the subspace constructed from 3 coded messages is 2.
Thus, it is sufficient for the node 1 to broadcast the 2 basis vectors and their linear combination coefficients.
Since it requires 2 linear combination coefficients in $\mathbb{F}_{2}$ for each coded message, the number of bits sent by the node 1 is $2\cdot\frac{T}{2}+3\times 2=T+6$, which is smaller than $\frac{3T}{2}$, the number of bits required to send the message in the original form, when $T$ is larger than 12\footnote{This is not the case for the example because $T=6$ is sufficient for representing the coded messages. However, $T$ would take very large values for big data applications.}.
This is because the dimension of the subspace constructed from the messages is smaller than the number of messages.
This example can be considered as a special case of the problem of word count in a large document.
In many cases, many words do not appear or appear only a small number of times in the divided blocks of the document.
In such cases, many intermediate values would have the same value, resulting in a smaller subspace rank.

\begin{figure}[t]
   \centering
   \includegraphics[keepaspectratio=true, width=\linewidth]{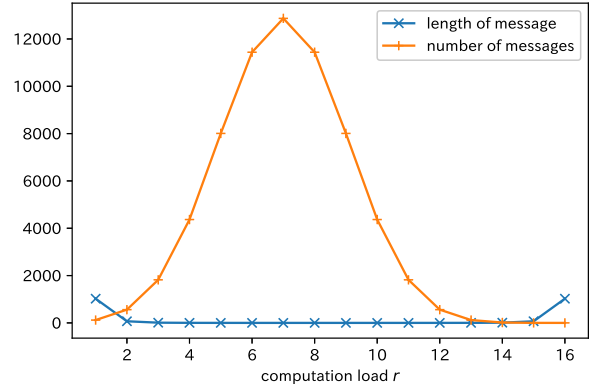}
   \caption{The length and the number of messages messages sent by a node for a problem of computing linear transforms, where the system is $K=Q=16, N=128$ and the parameters of the problem are $q=2, m=2048$. We can see that the length of the message is smaller than the number of messages when $2\le r\le 14$.}
   \label{fig:length_number}
\end{figure}

Next example is the problem to compute linear transforms of high-dimensional vectors, which is a critical step in several machine learning and signal processing applications.
Consider a problem to compute linear transforms in which given a matrix $\bm{A}\in\mathbb{F}_{q}^{m\times n}$ and $N$ input vectors $\bm{x}_{1},\ldots,\bm{x}_{N}\in\mathbb{F}_{q}^{n}$, we want to compute $\bm{y}_{1}=\bm{A}\bm{x}_{1},\ldots,\bm{y}_{N}=\bm{A}\bm{x}_{N}$.
We consider the problem to compute these linear transforms in a distributed computing system with $K$ nodes.
There are various ways to compute the linear transforms in a distributed manner.
One of such methods is that the matrix $A$ is divided by rows into submatirices $\left\{\bm{A}_{k}:k=1,\ldots,K\right\}$ and the linear functions defined by $\bm{A}_{k},k=1,\ldots,K$ are regarded as the reduce functions. 
Each server $k$ computes some of $\left\{A_{k}\bm{x}_{i}:i=1,\ldots,N\right\}$ and these output vectors are the intermediate values.
In this case, each intermediate value is represented by $T=m\log_{2}q/K$ bits.
Consider the case where $s=1$, that is, each reduce function is computed only once by a node.
Then, the length of the encoded messages is $\frac{mN\log_{2}q}{rK\binom{K}{r}}$ and the number of the messages sent by a node is $\binom{K}{r+1}$.
The rank of the subspace constructed from the encoded messages is smaller than or equal to $\min\left\{\frac{mN\log_{2}q}{rK\binom{K}{r}},\binom{K}{r+1}\right\}$ and it is smaller than the number of messages $\binom{K}{r+1}$ for some cases.
For a system with $K=Q=16, N=128$ and a problem with $q=2, m=2048$, the relation between the length and the number of the encoded messages sent by a node is depicted in Fig. \ref{fig:length_number}.

In some cases, the rank of the subspace is even smaller.
For example, as in \cite{dutta2016short}, one may construct the submatrices $\left\{\bm{A}_{k}:k=1,\ldots,K\right\}$ so that each submatrix is sparse.
Furthermore, in machine learning applications, vectors $\bm{x}_{1},\ldots,\bm{x}_{N}$ are often sparse \cite{duchi2011adaptive}.
Then, the intermediate values are zero vectors with high probability and they are linearly dependent.

Another example is also the problem to compute linear transforms of high-dimensional vectors, but in a situation that there would be some straggling nodes in the system.
For this problem, the scheme that combines the CDC scheme and error-correcting codes are proposed in \cite{li2016unified,zhang2019improved}.
In these schemes, error-correcting linear codes are applied to the matrix $\bm{A}$ and each server $k$ stores
\begin{align}
    \bm{U}_{k}=\bm{E}_{k}\bm{A}, 
\end{align}
where $\bm{E}_{1},\ldots,\bm{E}_{K}$ are designed so that the system can compute $\bm{A}\bm{x}$ even if some straggling nodes exist.
Each server $k$ compute some of $\left\{\bm{U}_{k}\bm{x}_{i}:i=1,\ldots,N\right\}$.
From the construction, $\bm{U}_{1},\ldots,\bm{U}_{K}$ are linearly dependent.
As a consequence, the intermediate values are also linearly dependent.

\subsection{Proposed Scheme}

The only difference of the proposed scheme from the scheme presented in section \ref{section2} is the way to construct the coded message in the shuffle phase.
First, each node compute the coded messages of the intermediate values in the same way as in (\ref{X_k_S}).
Each $\bm{X}_{k}^{\mathcal{S}}$ is a length-$\binom{|\mathcal{S}|-2}{r-1}\binom{r}{|\mathcal{S}|-s}\frac{\eta_{1}\eta_{2}T}{r}$ binary vector.
For $\max\left\{r+1,s\right\}\le \ell\le \min\left\{r+s,K\right\}$, there are a total of $\binom{K-1}{\ell-1}$ subsets of $\left\{1,\ldots,K\right\}$ with size $\ell$ that contain the node $k$.
Consider a set of messages 
\begin{align}
    V_{k,\ell}=\left\{\bm{X}_{k}^{\mathcal{S}}:\mathcal{S}\subseteq \left\{1,\ldots,K\right\}, k\in\mathcal{S}, |\mathcal{S}|=\ell\right\}.\label{subspace}
\end{align}
The set contains $\binom{K-1}{\ell-1}$ of length-$\binom{\ell-2}{r-1}\binom{r}{\ell-s}\frac{\eta_{1}\eta_{2}T}{r}$ vectors.
This set can be considered as a linear subspace.
Let $\rho_{k,\ell}$ be the rank of the subspace (\ref{subspace}).
The node $k$ computes the $\rho_{k,\ell}$ basis of the subspace and send the basis and linear combination coefficients which are needed to recover the original coded messages.
We call the proposed scheme \textit{CDC-LD}.
Since each node send $\binom{K-1}{\ell-1}$ coded messages and it requires $\rho_{k,\ell}$ linear combination coefficients in $\mathbb{F}_{2}$ to express each of them, the number of the bits sent by node $k$ is $\rho_{k,\ell}\binom{\ell-2}{r-1}\binom{r}{\ell-s}\frac{\eta_{1}\eta_{2}T}{r}+\rho_{k,\ell}\binom{K-1}{\ell-1}=\left(\binom{\ell-2}{r-1}\binom{r}{\ell-s}\frac{\eta_{1}\eta_{2}T}{r}+\binom{K-1}{\ell-1}\right)\rho_{k,\ell}$.
As $\ell$ varies from $\max\left\{r+1,s\right\}\le \ell\le \min\left\{r+s,K\right\}$, the proposed scheme has the following performance.
\begin{theorem}
The proposed scheme achieves the following communication load.
\begin{multline}
    L_{\mbox{CDC-LD}}\left(r,s, T, \left\{\rho_{\ell}\right\}\right)=\\
    \sum_{\ell=\max\left\{r+1,s\right\}}^{\min\left\{r+s,K\right\}}\left(\frac{\binom{\ell-2}{r-1}\binom{r}{\ell-s}}{r\binom{K}{r}}+\frac{K\binom{K-1}{\ell-1}}{QNT}\right)\rho_{\ell},
\end{multline}
where $\rho_{\ell}$ is defined as $\rho_{\ell}\triangleq \frac{\sum_{k=1}^{K}\rho_{k,\ell}}{K}$.

Especially, when $s=1$,
\begin{align}
     L_{\mbox{CDC-LD}}\left(r, T, \rho_{r+s}\right)=
    \left(\frac{1}{r\binom{K}{r}}+\frac{K\binom{K-1}{r}}{QNT}\right)\rho_{r+1}.
\end{align}
\end{theorem}

\begin{figure}[t]
   \centering
   \includegraphics[keepaspectratio=true, width=\linewidth]{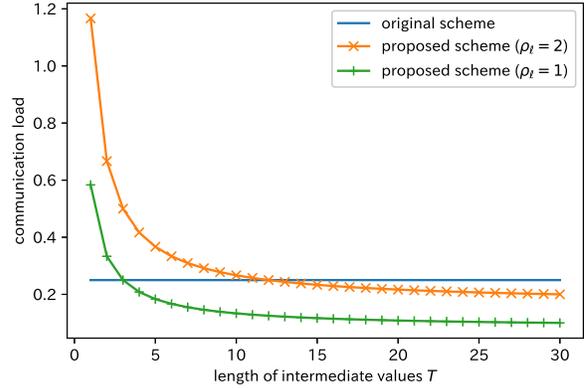}
   \caption{The communication loads of the original scheme and the proposed scheme as functions of the length of the intermediate values $T$, where the parameters of the system is $K=4, N=6, Q=4, r=2, s=1$. It does not depend on $T$ for the original scheme.}
   \label{fig:load_compare}
\end{figure}

The communication load of the proposed scheme depends on the values of $Q, T, \rho_{\ell}$, while the original scheme does not.
Therefore, it depends on these values whether the proposed scheme is more efficient than the original scheme.
The communication loads of the original scheme and the proposed scheme are depicted in Fig. \ref{fig:load_compare} for different values of $T$, where $K=4, N=6, Q=4, r=2, s=1$.
We can see that the communication load of the proposed system decreases as the length $T$ of the intermediate values increases.
In this case, the proposed scheme has better performance compared to the original scheme provided that $T\ge 12$ and the average rank $\rho_{\ell}\le 2$.

\begin{figure}[t]
   \centering
   \includegraphics[keepaspectratio=true, width=\linewidth]{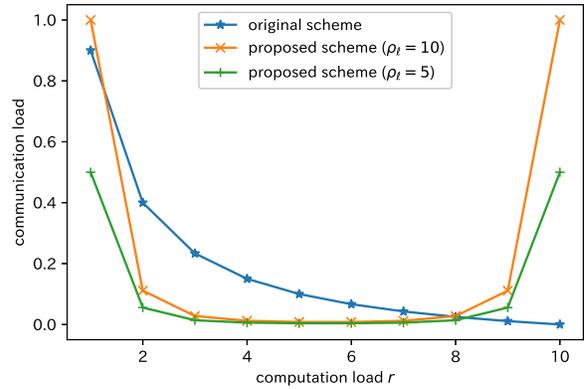}
   \caption{The computation-communication trade-off of the original scheme and the proposed scheme. The parameters of the system is K=10, N=2520, Q=360, T=64.}
   \label{fig:load_compare2}
\end{figure}

Fig. \ref{fig:load_compare2} shows the trade-off relationship between the computation load and the communication load for the system with $K=10, N=2520, Q=360, T=64$.
We can see that the computation load is a medium value, the proposed methods perform better than the original scheme.

\subsection{Discussion}

The proposed system utilizes the structure of the map functions and no assumptions are put for the reduce functions.
So we can combine the proposed method with the method proposed in \cite{li2018compressed} if the map functions have some linear dependent structure and the reduce functions are the linear aggregation where the reduce function is the sum of intermediate values.

Our proposed method can be considered as one of the methods of compressing the intermediate values.
Therefore, the lower bound of the communication load of each server to send its intermediate value is given by its entropy.
However, coding schemes with good compression performance often have large computational complexity for coding and decoding, resulting in an increase in the computational time of the overall system.
In our proposed scheme, it has to compute the basis of the subspace.
Although its computation cost is very high in some cases, there are some cases that it is not so high.
For example, we can choose standard basis for the problem of computing linear transforms.
In those cases, we can obtain a cost-effective improvement.

\section{Conclusion}

We have developed a new CDC scheme that improves the computation-communication trade-off by utilizing the linear dependence structure of the messages sent by a node during the data shuffling phase.
The central idea of the proposed scheme is that the messages constructed in the CDC scheme have linear dependency for some applications.
As far as the author knows, there has been research to improve the CDC scheme by using the properties of the reduce functions, but our research is the first attempt to improve the CDC scheme by using the property of the map functions.
While the proposed scheme requires an additional computation cost to compute the basis of the subspace, it can obtain a cost-effective improvement in some applications.
Analysing the degree of improvement for some specific applications, such as the word count problem or the problem of computing linear transforms, is for further study.


\section*{Acknowledgment}

This research is partially supported by the Kayamori Foundation of Informational Science Advancement and No. 19K12128 of Grant-in-Aid for Scientific Research Category (C) and No. 18H03642 of Grant-in-Aid for Scientific Research Category (A), Japan Society for the Promotion of Science.




\newpage

\bibliographystyle{IEEEtran}
\bibliography{ref}

\end{document}